\documentclass[prd,aps,preprintnumbers,twocolumn]{revtex4}
\usepackage{epsfig}
\usepackage{amsmath}
\def \be  {\begin{equation}}
\def \ee  {\end{equation}}
\def \ba  {\begin{eqnarray}}
\def \ea  {\end{eqnarray}}
\def \baa {\begin{eqnarray*}}
\def \eaa {\end{eqnarray*}}
\def \bb  {\begin{thebibliography}}  
\def \eb  {\end{thebibliography}}

\def \lab #1 {\label{#1}}

\def \fracs #1#2 {\mbox{\small $\frac{#1}{#2}$}}

\def \bin #1#2 {{\left({#1}\atop{#2}\right)}}
\def \as {\relax\ifmmode\alpha_s\else{$\alpha_s${ }}\fi}

\def \al #1 {\frac {\as({#1})}{\pi} }
\def \ds #1 {\ooalign{$\hfil/\hfil$\crcr$#1$}}

\newcommand \bea{\begin{eqnarray}}
\newcommand \eea{\end{eqnarray}}

\newcommand \vep {\varepsilon}

\def\hepph  #1 {{\tt hep-ph/#1}}

\begin{document}

\preprint{YITP-SB-07-21}
\preprint{BNL-NT-07/27}

\renewcommand{\thefigure}{\arabic{figure}}

\title{Color Transfer in Associated Heavy-Quarkonium Production}

\author{ Gouranga C.\ Nayak$^a$, Jian-Wei Qiu$^{b,c}$ and George Sterman$^a$}

\affiliation{{}$^a$C.N.\ Yang Institute for Theoretical Physics,
Stony Brook University,
Stony Brook, New York 11794-3840, USA}
\affiliation{{}$^b$Department of Physics and Astronomy,
Iowa State University,
Ames, Iowa 50011-3160, USA}
\affiliation{{}$^c$Physics Department, Brookhaven National Laboratory, Upton, NY 11973, USA}
\date{\today}

\begin{abstract}
We study the production of heavy quarkonium
in association with an additional heavy pair.
We argue that important contributions
may come from phase space regions
where three heavy fermions are separated by
relative velocities much lower than the speed of 
light, and to which standard effective field theories do not
apply.  In this region, infrared sensitive
color exchange is specific to the presence of the unpaired (anti)quark.
This effect vanishes as the motion of the additional particle 
becomes relativistic
with respect to the pair, and is completely absent for 
massless quarks and gluons in the final state.  
\end{abstract}

\maketitle

In heavy-quarkonium production, the formation of
the heavy-quark pair can be perturbative.
The subsequent evolution of the quark pair
is conventionally treated in the language of
effective theories, 
most notably non-relativistic QCD (NRQCD) \cite{bodwin94}.
In NRQCD, non-perturbative dynamics
is organized through 
a factorization characterized by
a joint expansion in the strong coupling,
$\alpha_s$ and $v$, the relative velocity.
This expansion is well-justified for heavy-quarkonium
decays, although it still lacks a  compelling 
proof for production processes.
It is clear, however, that NRQCD cannot
apply directly to reactions in which three
heavy particles, two quarks and an antiquark for
example, are produced close together in phase 
space, simply because there is no unique choice
of relative velocity $v$.   

Although the application of NRQCD to
production processes has had many successes \cite{cdf,QWGBrambilla},
there remain well-known data anomalies,
particularly in connection with  the polarization observations
at the Tevatron \cite{cdfpolar} and 
the size of the associated production cross section 
for $J/\psi$ with $c\bar c$ pairs, 
as seen by the BELLE and Babar collaborations 
\cite{associated}.   
Regarding the latter, the rate for $J/\psi$ associated production
  is unexpectedly large compared to 
  $J/\psi$ with light hadrons.
The importance of associated production within the
formalism of NRQCD has
recently been explored at leading order (LO)  
for hadroproduction in Ref.\ \cite{Artoisenet:2007xi} and
at next-to-leading order (NLO)  for
$\rm e^+e^-$ annihilation in Ref.\ \cite{Zhang:2006ay}.
We will see, however, that NRQCD may not be the
complete story for these processes.

Associated pair production
for $J/\psi$ and related quarkonium states
provides access to
a novel kinematic region, where the
heavy quarkonium is produced close in phase
space to open heavy flavor.  The presence of
the additional quark, itself a source of color, could 
influence hadronization.  In addition, studying this configuration sheds light
on NRQCD factorization \cite{bodwin03}
for processes without associated heavy quark
pairs, and we will make contact with
our previous work on that important question \cite{Nayak:2005rt,Nayak:2006fm}.
We begin with a discussion
of matching and infrared poles in dimensional regularization 
for NLO corrections to
associated production of heavy quarks $Q$, of mass $m$, 
in $\rm e^+e^-$ annihilation.

\begin{figure}[t]
\begin{center}
\begin{minipage}[c]{1in}
\epsfig{figure=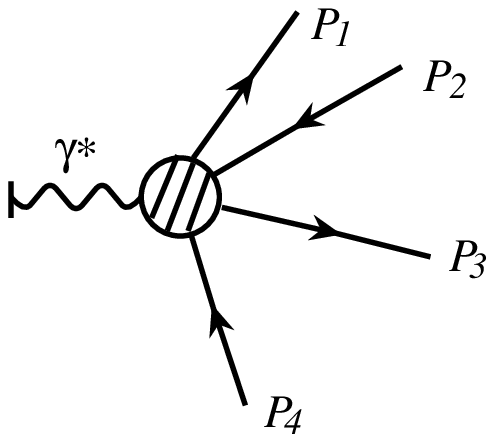,width=1in}
\end{minipage}
\hskip 0.2in
=
\hskip 0.2in
\begin{minipage}[c]{1in}
\epsfig{figure=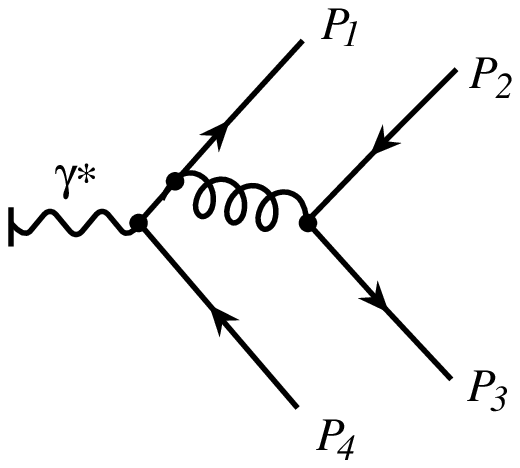,width=1in}
\end{minipage}
\hskip 0.2in
+
\hskip 0.2in
\begin{minipage}[c]{1in}
\epsfig{figure=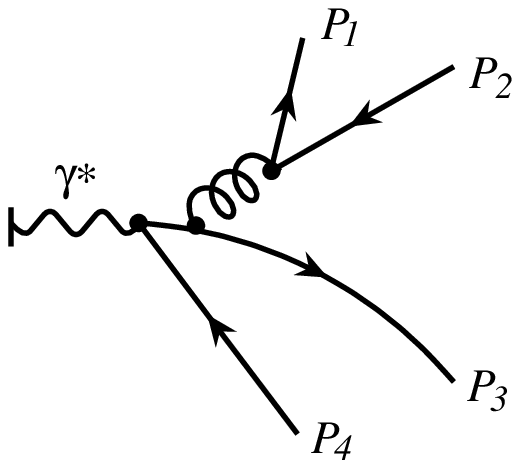,width=1in}
\end{minipage}
\hskip 0.2in
+\ $\cdots$

\caption{Representative lowest order diagrams for
a virtual photon decaying into two heavy quark pairs, for
fixed quark momentum  $P_1$ and antiquark momentum $P_2$.
The pair $P_1$ and $P_2$ will be associated with a bound state.
}
\label{lofig}
\end{center}
\end{figure}

Diagrams for ${\rm e^+e^-} \to Q\bar Q+Q\bar Q$
are shown in Fig.\ \ref{lofig}. 
Because 
we will study final states with two pairs, we will
find it useful to refer to the pair that forms the bound
state as the ``active" pair.
To be specific, we will
denote the quark and antiquark momenta of the
active pair as $P_1$ and $P_2$, respectively,
with total and relative momenta 
and rest-frame relative velocity given by
\bea
P_1=\frac{P}{2}+q,\;  P_2 = \frac{P}{2}-q,\; v^2=\frac{\vec{q}\, {}^2}{E^*{}^2}
= 1-\frac{4m^2}{P^2}\, ,
\label{Pqvdefs}
\eea
where $2E^*$ is the total center of mass energy of the active
pair, and where in the
second expression for $v^2$ we give the invariant form.
In the diagrams of Fig.\ \ref{lofig},  we refer to the 
additional heavy particle that is closest in phase
space to the active pair as the ``spectator"
quark (or antiquark).   

The LO NRQCD cross sections for
the amplitudes of Fig.\ \ref{lofig}
are given by
\begin{equation}
d\sigma_{e^+e^-\to H+X}(p_H) = \sum_n\; 
d\hat\sigma_{e^+e^-\to  Q\bar{Q}[n]+X}(p_H)\,
\langle {\mathcal  O}^H_n\rangle\, ,
\label{nrfact}
\end{equation}
where $d\hat\sigma_{e^+e^-\to Q\bar{Q}[n]
+X}(p_H)$ is an NRQCD coefficient function, 
projected onto heavy quark pair state $Q\bar Q[n]$.
The corresponding NRQCD operators are of the form
  \cite{bodwin94} 
\ba
{\mathcal O}^H_n(0)
&=&
\sum_N\ \chi^\dagger(0){\mathcal \kappa}_n\psi(0)\, \left | N,H\right\rangle\,
\left\langle N,H \right|\, \psi^\dagger(0){\mathcal \kappa}'_n\chi(0)
\, ,
\nonumber\\
\label{Ondef1}
\ea
with $\kappa_n$ and $\kappa'_n$
projections for color (octet or singlet) and spin, and where
we suppress gauge links \cite{Nayak:2005rt}.
At LO, NRQCD singlet configurations are 
present only in diagrams where the active quark and antiquark arise from
different lines, as in the first graph on the right-hand side 
of Fig.\ \ref{lofig}.
In the second graph, the active pair is purely octet,
and is expected to contribute at a much lower level to the cross section.

For NRQCD factorization to be predictive, it is necessary that 
the coefficient function, $d\hat\sigma_{e^+e^-\to 
Q\bar{Q}[n]
+X}(p_H)$ be infrared safe at higher orders.
Not only infrared singularities, but
also singular dependence in any of the relative
velocities must be absorbed into the long-distance matrix elements
of the effective theory,
a process referred to as matching.
At LO, all gluons are
off-shell by at least a multiple of $m$,
and the full LO amplitudes may be absorbed into coefficient functions.
Nontrivial issues of factorization and
matching first arise at NLO.

It is a major undertaking to compute the full NLO corrections to
the singlet NRQCD cross section \cite{Zhang:2006ay},
and all the more so for the full four-particle
cross section at finite relative velocities.  Nevertheless, it is relatively
straightforward to check the self-consistency of the NRQCD 
factorization at NLO by verifying the cancellation of infrared poles.  

\begin{figure}[h]
\begin{center}
\begin{minipage}[c]{1in}
\epsfig{figure=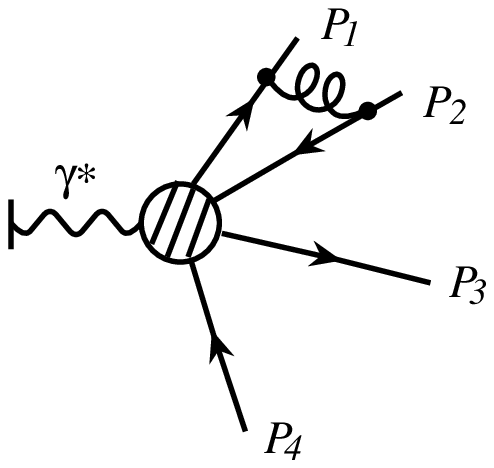,width=1in}
\end{minipage}
\hskip 0.2in
+
\hskip 0.2in
\begin{minipage}[c]{1in}
\epsfig{figure=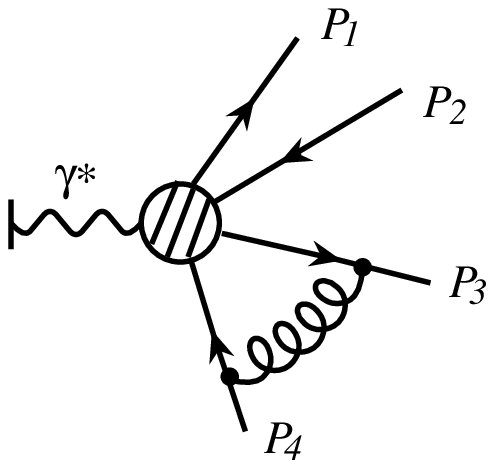,width=1in}
\end{minipage}

+
\hskip 0.2in
\begin{minipage}[c]{1in}
\epsfig{figure=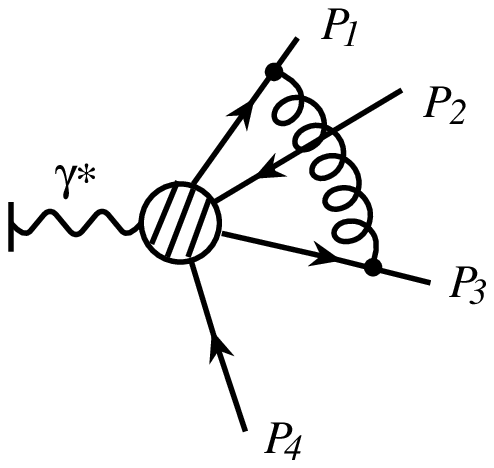,width=1in}
\end{minipage}
+
\hskip 0.2in
\begin{minipage}[c]{1in}
\epsfig{figure=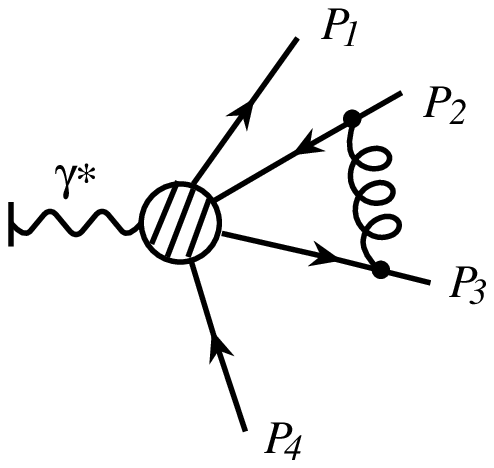,width=1in}
\end{minipage}
\hskip 0.2in
+ $\cdots$

\caption{Diagrams for one-loop virtual infrared corrections, 
where the blob represents a complete set of lowest order
Feynman diagrams, as shown in Fig.~\protect\ref{lofig}.
}
\label{nlofig}
\end{center}
\end{figure}

As noted above, when more than two
relative velocities are small, the standard
NRQCD power counting does not apply directly, simply because
there are now three dimensionless scales, 
$\beta_{ij}\equiv \sqrt{1-4m^2/(P_i+P_j)^2}$,
$i\ne j=1,2,3$,  instead of the single $v=\beta_{12}$ 
defined by the active pair.
Thus, where there was once one ``soft" region $mv$ and
one ``ultrasoft" region $mv^2$, there are now three of each.
Let us see how these considerations emerge in the NLO diagrams
of Fig.\ \ref{nlofig}.  

We stress that the region of
small relative velocity for three of the four particles in
double heavy-pair production is leading-power in
the overall center-of-mass energy, $\sqrt{s}$.  
For example, the
two diagrams on the right-hand side of Fig.\ \ref{lofig} 
contribute to the fragmentation of a heavy quark 
into the quark plus a pair.  
Such contributions also occur as part of the high-$p_T$ 
cross section for associated production in hadronic collisions,
as studied in Ref.\ \cite{Artoisenet:2007xi}.  It is worth noting as well
in this context that in such  fragmentation-like contributions,
the natural scale for $\alpha_s$ is $m$, while for the corresponding LO
contributions in which a light pair recoils
from a pair of heavy quarks, the natural scale for $\alpha_s$
is $\sqrt{s}$.  This is an additional source of relative enhancement
for associated production, compared to $J/\psi$ 
cross sections with light quanta only.

As discussed
extensively in \cite{Nayak:2005rt}, factorization requires the cancellation
of all infrared gluons that are not ``topologically factorized".   
Only such contributions will match the single-pair matrix elements
of NRQCD.
We may think of a topologically factorized singularity 
as one that can either be absorbed into
a quarkonium matrix element or that does not involve either of the active
heavy quarks.  An example of the
former is the first diagram in Fig.\ \ref{nlofig}, and of the
latter, the second.   After a sum over real
gluon emission at fixed color for the active quark pair, 
exchanges between spectators cancel.

The third and fourth diagrams of Fig.\ \ref{nlofig}, however,
are not topologically factorized, and include the transfer
of color between the active pair 
produced at short distances and the spectator.
A pair produced as an octet in the
hard scattering may end up as a singlet in the final state.
Let us see how this mechanism enters the NLO corrections, and why it
is not a problem except in the region where 
the spectator comes close to the active pair.

When $ \beta_{ij}\ll 1 $,
the infrared pole behavior of any of the exchanges
shown in Fig.\ \ref{nlofig} is given up to a sign by
\begin{widetext}
\bea
&&  -\, i\,g^2\,  \int \frac{d^D k}{(2\pi)^D}\
\frac{4P_i\cdot P_j}
     {[2P_i\cdot k +k^2+ i\epsilon]
      [-2P_j\cdot k +k^2+ i\epsilon]
      [k^2 + i\epsilon]}\, 
=  \frac{\alpha_s}{2\pi} \left[-\, \frac{1}{2\varepsilon}
\left(\frac{1}{ \beta_{ij}} +\beta_{ij}\, \right)
\left(  2\beta_{ij} -i \pi   \right)\ 
+ \dots\right],
\label{nloij}
\eea
\end{widetext}
with corrections that are finite as $\varepsilon=2-D/2\rightarrow 0$ or
that vanish as higher powers of $\beta_{ij}$.   We will concentrate on
the imaginary term, which is singular 
for both vanishing $\beta$ and $\varepsilon$.
Consider now the  imaginary contributions of Eq.\ (\ref{nloij}) to
the NLO ampliude ${\cal A}$,
traced over the colors of the active pair to enforce
a singlet configuration in the final state.
Denoting by ${\cal A}_{ij}$ the amplitude with
an exchange between $P_i$ and $P_j$, we have, 
%suppressing color factors,
suppressing common color factors in the quark and antiquark couplings,
\ba
\hspace{-3mm}
{\rm Im}
\left[{\cal A}_{13}+{\cal A}_{23}\right]
&=& \frac{\as}{4 \vep}\,  {\cal A}^{(0)}(P_i)
\left[
\frac{1+ \beta^2_{13}  }{ \beta_{13}}
-
\frac{1+ \beta^2_{23}}{ \beta_{23}}
\right],
\label{allorder}
\ea
where ${\cal A}^{(0)}$ represents the LO amplitude.  
From this soft gluon correction, the third 
and fourth diagrams in Fig.\ \ref{nlofig} inherit color singlet configurations
for the pair $P_1,P_2$ in the final state at NLO.  We should mention that 
this contribution is subleading in the number of colors $N_c$.
Indeed, it constitutes an alternative mechanism to the ``preconfinement"
scenario, identified long ago in Ref.\ \cite{Amati:1979fg}, and built into 
most hadronization models in event generators \cite{Ellis:1991qj}.

The essential feature of Eq\ (\ref{allorder}) is the
presence of poles in the relative velocities $\beta_{13}$ and
$\beta_{23}$.   These poles cancel
identically for $P_1\cdot P_3=P_2\cdot P_3$
or for $P_3^2=0$.  In the former case, the relative
velocity of the spectator quark and the active quark
equals the relative velocity of the spectator and
the active antiquark.  
 Here, soft gluons
emitted by the spectator cannot resolve the
charges of a perfectly co-moving pair
in a singlet color state.
Correspondingly, when the spectator is 
lightlike, its relative velocity to either active line is unity,
and the two terms cancel.  
Thus for pair production in the presence of
massless quanta only (light quark pairs and gluons)
the color transfer mechanism is completely absent.
 This is an essential ingredient in the
verification of NRQCD factorization at NNLO 
in Refs.\ \cite{Nayak:2005rt,Nayak:2006fm}.
Evidently,  a light-like spectator can neither
resolve {\it nor produce} a color singlet even when the
quark and antiquark are not co-moving.  
We see here, however, that such a cancellation
requires that any heavy spectator be 
moving relativistically with respect to the active pair.
For nonrelativistic motion, there is a nontrivial
color transfer from the triplet spectator quark,
that catalyzes the transition between 
octet and singlet.

If, as in Ref.\ \cite{Zhang:2006ay},
the active quark relative velocity $v$ appears only
as a regulator for active-active gluon exchange,
nonfactorizing singularities never occur.
At the same time, at NLO, this 
cancellation is not exact in the low-velocity
region for the spectator.    The color transfer contribution is
neither topologically factorized nor infrared finite,
and cannot be absorbed into the coefficient function
for the singlet production amplitude. 
Although this effect
is purely imaginary at NLO, it will have a physical
role at higher orders.

Since the infrared 
correction to the amplitude, ${\cal A}$ of (\ref{allorder})
vanishes at vanishing active-pair
relative momentum, it is natural to expand
${\cal A}$ in $v$ at fixed relative pair-spectator velocity,
which we will refer to as $\beta_S$, between
the spectator quark (of momentum $P_3$)
and $P/2$ \cite{nqsinprep},
\bea
\beta_S = \sqrt{\frac{-q_S^2}{m^2-q_S^2}}, \quad
P_3^\mu = \frac{P_0^\mu}{2}\sqrt{1-\frac{q_S^2}{m^2}}+q_S^\mu,
\eea
with $P_0^\mu=(2m,0)$ and $q_S\cdot P_0=0$.
Expanding in the relative
momentum of the active pair in the spirit of NRQCD,
we find
\ba
\frac{1}{ \beta_{13}}
-\frac{1}{ \beta_{23}} \sim
 -\frac{4}{ \beta^3_S}\ \frac{q_S\cdot q}{m^2}
\sim
 \frac{4}{\beta^2_S}\ v\, \cos\phi_S\, ,
\label{firstorder}
\ea
where in the second relation $\phi_S$ is the angle 
between the vectors $\vec q_S$ and $\vec q$.  The presence of the
cosine, which is odd under the interchange of 
the members of the active pair, is an expression
of the dipole nature of this contribution to  the amplitude.

Although the expansion in Eq.\ (\ref{firstorder}) is
applicable in only a limited region of phase space,
it is a region in which the active pair and spectator
quark form  a system of invariant mass of order $(3m)^2$.
In this kinematic region, the squares of NLO corrections
to LO diagrams  
like those on the right-hand side of Fig.\ \ref{lofig}
contribute to the cross section at leading power
in 
the overall center-of-mass
energy.  At NNLO, the squares of the imaginary poles of  Eq.\ (\ref{nloij})
produce $\alpha_s^4(1/\varepsilon^2)$
corrections to the cross section.  All other
diagrams contribute only at the single-pole level \cite{nqsinprep}.
We note again, that in accordance with
the results of Ref.\ \cite{Nayak:2005rt}, this contribution vanishes
in the limit $m\rightarrow 0$, and so is consistent
with NRQCD factorization at large $p_T$.

In summary, we have identified a new, leading-power ``color-transfer"
infrared-divergent contribution to the formation of
color-singlet pairs, which is specific to 
associated production.   This contribution must be added to that of the
familiar color-octet mechanism \cite{bodwin94}.   
The infrared divergence signals that color transfer cannot be absorbed into
a coefficient function, while its dependence on $\beta_S$ signals
that it cannot be absorbed into matrix elements in an effective
field theory  based on the dynamics of a single pair.   
This mechanism is non-leading in
the number of colors, although that by itself
should not eliminate it from consideration.

It is natural to try and estimate the  
magnitude of this new effect.   A plausible estimate can be
given by elaborating on the analogy
just observed between the normal color octet
mechanism and color transfer. 
In both cases, an S-wave pair is transformed to
P-wave by the gluonic
field strength contracted with the pair's color dipole.
In the octet mechanism, the field couples to the
polarization of an emitted gluon, while
in color transfer it is coupled to the Coulomb field of the spectator.
To return to an S-wave
state, we will assume a $\beta_S$-independent $v^2$ suppression,
following NRQCD power counting \cite{bodwin94,QWGBrambilla}.
%%
%%In  the former, an S-wave quark pair essentially exchanges color
%%with the vacuum to become a singlet, while in the latter, 
%%the color is transfered to
%%a spectator.  
%%Both are color electric dipole  transitions, 
%%producing $P$ wave intermediate states,
%%and  are thus proportional  to $v^2$ in NNLO perturbative
%%cross sections,
%%and are suppressed by an overall $v^4$ for production of 
%%S-wave quarkonia like $J/\psi$ \cite{bodwin94,QWGBrambilla}.
At fixed $\beta_S$, let us
assume that the nonperturbative formation  of  quarkonium  state, $H$
is similar for these two transitions.
We then square the $\beta_S$-dependent kinematic factor of
Eq.\ (\ref{firstorder}), and replace the $v$-dependence by the
 NRQCD matrix element
$\langle{}^3{\mathcal S}_8^H\rangle$.   In the total
cross section at fixed $\beta_S$,  such a quantity would 
be additive to  the color singlet contribution, which dominates
in the NRQCD treatment.  We thus have, schematically,
\begin{widetext}
\bea
d\sigma^{\rm tot}_{e^+e^-\to H+X}(p_H) 
\sim d\hat\sigma_{e^+e^-\to 
Q\bar{Q}
[S_1]+Q'(\beta_S)}(p_H)\, 
 \langle{}^3{\mathcal S}_1^H \rangle
+
 d\hat\sigma_{e^+e^-\to  Q\bar{Q}
[S_8]+Q'(\beta_S)}(p_H)\, 
\frac{ \langle{}^3{\mathcal S}_8^H \rangle}{\beta_S^4}\, ,
\label{colortransfersigma}
\eea
\end{widetext}
that is, the standard color singlet, plus a color-octet contribution
enhanced by the factor
$1/{\beta_S^4}$, with $\beta_S$ the velocity of 
the closest heavy spectator, $Q'=Q,\bar{Q}$.   
By numerical evaluation, we have 
verified \cite{nqsinprep} that the short-distance cross section,
$ d\hat\sigma_{e^+e^-\to  Q\bar{Q}
[S_8]+Q'(\beta_S)}(p_H)$
is also larger than the  corresponding singlet contribution.
At $\beta_S\sim 0.3$, our
estimate of color transfer 
would then become
competitive with the color singlet cross section at the same $\beta_S$.   
Of course, 
 for $\beta_S<v$ an expansion in $v/\beta_S$
loses meaning,
and we do not offer Eq.\ (\ref{colortransfersigma}) as a realistic estimate, 
especially for $J/\psi$, with $v\sim 0.5$.  We do present it, however, as an
indication of the possible importance of the new mechanism.

A signature of the color-transfer process 
would clearly be an enhancement in any open heavy flavor
at low relative velocity to quarkonia,
 in hadronic scattering as well
as leptonic annihilation.
Another
sign could be in studies
of polarization.  As observed in Ref.\ \cite{Artoisenet:2007xi} in the
context of LO NRQCD, associated production 
leads to a more complex polarization structure than
that expected from a pure
octet mechanism.  Although color transfer does not
fall cleanly into the analysis of NRQCD, we might
expect much the same spin symmetry to apply,
so that a study of polarization effects 
can also help characterize this mechanism.

To develop an effective theory
that  incorporates color transfer,
we shall have to add new operators.
As already found in Ref.\ \cite{Nayak:2005rt} in the context
of NRQCD factorization, the long-distance
transition from octet to color  singlet states
can be associated with a nonlocal operator of the form
$ \int_0^\infty d\lambda' \,  \lambda' \,
   \left[\, P^\mu q^\nu F_{\nu\mu,a}(\lambda' P)\, \right]$,
   with $P$ and $q$ as in Eq.\ (\ref{Pqvdefs}).
   Noting that $\lambda' q$ can be interpreted as
   the configuration-space separation of a pair,
we see that this is the same operator that represents
 the singlet-octet transition in the
potential NRQCD Hamiltonian \cite{Brambilla:1999xf,Hoang:2002ae},
here extended to production.  It is clear,
however, that this correspondence holds only
in the region where we can strongly order the velocities.
More generally, we shall have to introduce 
products of three fields, as $\psi^\dagger_{P_1}\chi_{P_2}\psi^\dagger_{P_3}$,
whose interactions with the soft gluon field 
will be labeled by momenta (as in NRQCD) 
according to heavy quark effective theory.
In general the nonperturbative dynamics these operators will 
overlap with those of NRQCD,
as well as those of potential NRQCD.  It will require 
some care to avoid double counting and to
develop a formalism that can be matched to
the phenomenology of associated production.

We thank Geoff Bodwin for many useful discussions on factorization.
This work
was supported in part by the National Science Foundation, grants
PHY-0354776 and PHY-0354822, by the US
Department of Energy under Grants No.~DE-FG02-87ER-40371
and  DE-AC02-98CH-10886 and in part by the Argonne University of 
Chicago Joint Theory Institute (JTI)
Grant 03921-07-137.

\end{document}